\documentclass[]{spie}  
\usepackage[dvips]{graphicx}

\newcommand{\cmsq}{\hbox{cm$^{-2}$}}

\newcommand{\flux}{\hbox{erg~cm$^{-2}$~s$^{-1}$}}
\newcommand{\lumin}{{erg~s$^{-1}$}}

\newcommand{\msun}{\hbox{${M}_{\odot}$}}

\newcommand{\simgt}{\lower 2pt \hbox{$\, \buildrel {\scriptstyle >}\over {\scriptstyle\sim}\,$}}
\newcommand{\simlt}{\lower 2pt \hbox{$\, \buildrel {\scriptstyle <}\over {\scriptstyle\sim}\,$}}
\newcommand{\arcsec}{\mbox{$^{\prime\prime}$}}

\newcommand\farcs{\mbox{$.\!\!^{\prime\prime}$}}%
\newcommand{\asca}{{\emph{ASCA}}}
\newcommand{\xmm}{{\emph{XMM-Newton}}}
\newcommand{\chandra}{{\emph{Chandra}}}

\newcommand{\sax}{{\emph{Beppo-SAX}}}
\newcommand{\rosat}{{\emph{ROSAT}}}
\newcommand{\xeus}{{\emph{XEUS}}}

\newcommand{\ngst}{{\emph{NGST}}}
\newcommand{\genx}{{\emph{Generation-X}}}
\newcommand{\conx}{{\emph{Constellation-X}}}
\newcommand{\einstein}{{\emph{Einstein}}}
\newcommand{\heao}{\hbox{\emph{HEAO-1}}}

\newcommand{\eflux}{\mbox{photons~cm$^{-2}$~s$^{-1}$~keV$^{-1}$}}

\title{The Challenge to Large Optical Telescopes \\ from X-ray Astronomy}

\author{Ann  E. Hornschemeier\supit{a} 
\skiplinehalf
\supit{a}Penn State University, 525 Davey Lab, University Park, PA, USA \\
}

\authorinfo{Further author information: (Send correspondence to A.E.H.)\\: E-mail: annh@astro.psu.edu, Telephone: 1 814 863 0182\\ Address: 525 Davey Lab, University Park, PA 16802, USA}

\begin{document} 
  \maketitle 

\begin{abstract}

In the \rosat\ era of the mid-1990's, the problems 
facing deep X-ray surveys could be largely solved 
with 10 m class telescopes.  In the first decade 
of this new millennium, with X-ray telescopes 
such as the \chandra\ X-ray Observatory and 
\xmm\ in operation, deep X-ray surveys 
are challenging 10 m telescopes.  For example, 
in the Chandra Deep Field surveys, $\approx30$\% of the
X-ray sources have optical counterparts fainter than $R=25$ ($I=24$).

This paper reviews current progress with 
6--10 m class telescopes in following up sources 
discovered in deep X-ray surveys, including
results from several X-ray surveys which have depended on telescopes 
such as Keck, VLT and HET.   Topics 
include the prospects for detecting extreme 
redshift ($z > 6$) quasars and the  first 
detections of normal and starburst galaxies 
at cosmologically interesting distances in the X-ray band.

X-ray astronomy can significantly bolster the science case
for the next generation of large aperture (30--100~m) ground-based
telescopes and has already provided targets for these large telescopes
through the \chandra\ and \xmm\ surveys.  The next generation of
X-ray telescopes will continue to challenge large optical telescopes;
this review concludes with a discussion of prospects from new
X-ray missions coming into operation on a 5--30 year timescale.  

\end{abstract}



\section{INTRODUCTION}
\label{sect:intro}  

X-ray astronomy has made great strides in the past several years with
the larger effective areas and greatly improved spatial resolutions of
\chandra\cite{Weisskopf02} and \xmm\cite{Jansen01} (both launched in
1999).  However, the science results from deep X-ray surveys even
before these missions relied heavily on the information obtained from
large ground-based optical telescopes.  For example, those who carried
out the optical follow-up of the \rosat\ Ultra Deep Survey (UDS)
sources\cite{Hasinger98,Schmidt98,Lehmann01} broke through major
barriers only when 10~m telescopes became available.

The purpose of this contribution is to describe the science coming from
deep X-ray surveys, including the nature of the first supermassive
black holes in the Universe and the evolution of the high-energy emission
from normal and starburst galaxies since $z\approx1$.  These scientific
results have relied critically on ground-based optical follow-up with
large telescopes such as Keck, VLT, and the HET, fitting well with
this conference's theme of ``science with 6--10~m class
telescopes."

\subsection{An X-ray Astronomy Primer}

X-ray astronomy utilizes much smaller aperture
optics than in the optical band 
(the Wolter-type \chandra\ mirrors have diameters 
between 0.6--1.2~m and effective area $\approx100$--600~cm$^{2}$ over
0.5--8~keV), but 
these ``small" X-ray telescopes are revolutionizing our
understanding of the high-energy Universe.
This section covers some general terms and definitions in X-ray
astronomy for those who observe outside the X-ray band.

The X-ray surveys discussed here cover the 0.5--10.0~keV band, a
dynamic range of frequency as large as that from the near-infrared to
the ultraviolet. There is generally a division made between the ``soft"
and ``hard" X-ray bands, largely due to the nature of X-ray detectors.
The soft band is typically \hbox{ $0.25 \simlt  E \simlt 2$~keV } and
the hard band is \hbox{ $2 \simlt  E \simlt  10$~keV }.  Many
non-thermal X-ray sources are described empirically as a power law of
the form $F_{\rm E}=N_{\rm E}E^{-\Gamma}$ where $N_{\rm E}$ is the
normalization at 1~keV in units of \eflux. The parameter $\Gamma$ is
called the photon index.   The
shape of a source's X-ray spectrum is often described by its
``hardness": an object with a flat X-ray spectrum (smaller $\Gamma$) 
is referred to as ``hard".

X-rays are an important way to study supermassive black holes as they
are thought to be a universal property of accretion.  Throughout this
paper ``accreting supermassive black holes" is synonymous with ``active
galactic nucleus" (AGN).  Energetically X-rays are significant,
comprising between 2--$20\%$ of the radiant power of AGN\cite{Elvis94}.
The AGN X-ray emission at energies up to $\sim100$~keV is believed to
originate as ultraviolet and optical photons from the accretion disk
that illuminate a hot ($\sim10^{7-9}$~K) corona.  The corona, or
ionized plasma surrounding the black hole, reprocesses this radiation,
perhaps via inverse Compton scattering, and results in a non-thermal
X-ray spectrum\cite{Krolik99} (a power-law).  Though there is
significant scatter in its value, $\Gamma$ typically ranges from
1.7--2.3\cite{ReTu2000,GeoEtal2000} for radio-quiet luminous AGN.

Radio-quiet luminous AGN have X-ray luminosities in the range
$10^{42}-10^{45}$~\lumin (0.5--8~keV);  below this range AGN are
referred to as low-luminosity AGN (LLAGN), as they have much lower
luminosity than expected for accretion via a geometrically thin,
optically thick accretion disk onto a supermassive black
hole\cite{Shakura73}.  AGN are generally classified, both in the
optical and X-ray bands,
 by the level of obscuration toward the accretion activity.  The
classification of ``obscured/unobscured" refers to the X-ray spectral
classification of an AGN's spectrum, where the obscuration is described
by the hydrogen column density, $N_{H}$.  Optical classification is
split into ``Type I" and ``Type II" where Type I objects demonstrate
broad ($>1000$~km~s$^{-1}$) emission lines and Type II objects only
show narrow lines. Generally, X-ray obscured AGN are also Type II
objects, although obscuration in the X-ray is not necessarily
concordant with obscuration in the optical\cite{Gallagher02}.

The ``normal" galaxies discussed here are much less X-ray luminous than
typical AGN; the most luminous normal galaxies have
 $L_{\rm X} \approx 10^{42}$~\lumin ~(typical X-ray
 luminosities\cite{Fabbiano89} are
$L_{X}\approx10^{39}$--10$^{40}$~\lumin).  X-ray emission is an
important way to study star formation in galaxies, revealing 
stellar endpoints (i.e., compact objects in binaries, supernovae)
that are often much less visible in other wavebands.  Significant
amounts of matter may be found in the hot phase of the 
interstellar media (ISM) of galaxies, representing another key 
physical property of quiescent galaxies that requires X-ray observation
to be accurately measured.

While the dominant X-ray emission from AGN is generally confined to the
central part of the galaxy, the X-ray emission from normal galaxies is
more spatially extended\cite{Fabbiano92} as it arises from
accreting binary star systems, supernova remnants, and hot ISM. 
An LLAGN component may also be present, although these rarely dominate the
X-ray emission (LLAGN may have X-ray luminosities as low\cite{Ho01} as
$\approx 10^{38}$--$10^{39}$~\lumin).

X-rays break degeneracies concerning the origin of
far-infrared/submillimeter emission from high-redshift dusty
starbursts\cite{Bautz00,Barger01,davoXI}, providing another independent
diagnostic of the evolution of star-formation processes over cosmic
time.  X-rays are also more penetrating than ultraviolet emission which
has been used to diagnose levels of current
star-formation\cite{Madau96}.

\subsection{Deep X-ray Surveys and the X-ray Background}
\label{sect:xraysurveys}


   \begin{figure}
   \begin{center}
   \begin{tabular}{c}
   \includegraphics[height=10cm]{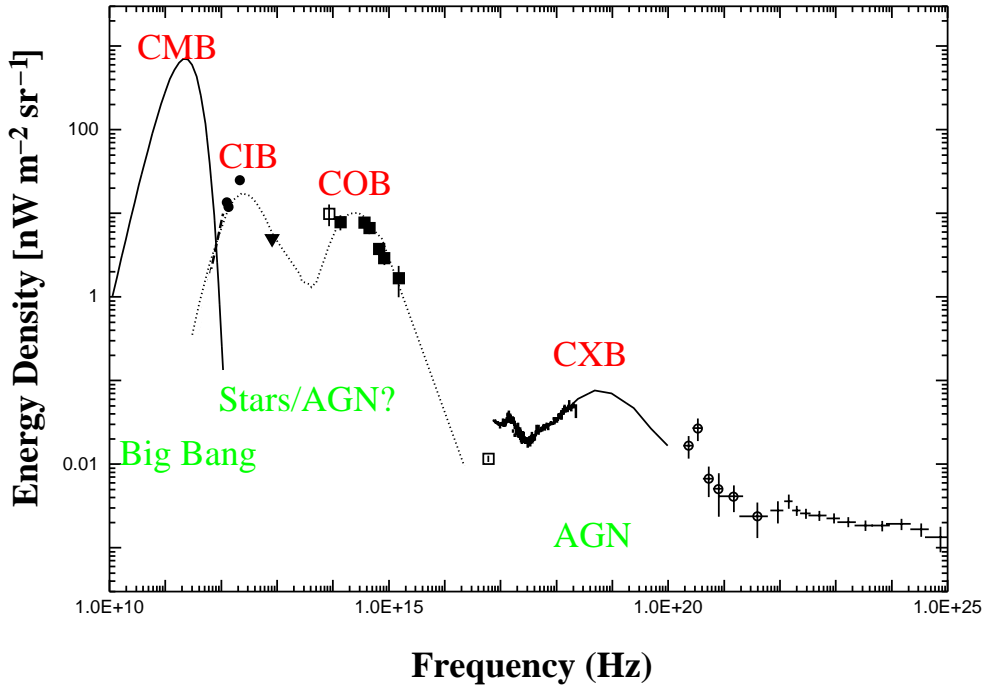}
   \end{tabular}
   \end{center}
   \caption[example]
   { \label{fig:EBL}
The cosmic energy density spectrum from
radio waves to high-energy gamma rays in a $\nu I_{\nu}$
representation.  This figure is adapted from 
Hasinger \& Gilli\cite{Hasinger00} with permission; 
the references for the data are contained within this article.  
The Cosmic Microwave Background (CMB) clearly dominates
the EBR. The other three distinct components are the Cosmic 
Far-Infrared Background (CIB), the
Cosmic Optical/Ultraviolet Background (COB), and the Cosmic
X-ray/Gamma-Ray Background (CXB).  
}
   \end{figure}

The imprint of every physical process over the history of the Universe
is present in the extragalactic background radiation
(EBR; for a review, see Hasinger \& Gilli\cite{Hasinger00}).
The spectrum of the EBR is shown in Figure~\ref{fig:EBL} and
is clearly dominated by the Cosmic Microwave Background\cite{Penzias65} (CMB),
which is the signature of the very early Universe at the epoch where 
matter and radiation decouple\cite{Mather94}.  At far-infrared
wavelengths, we are likely observing the re-radiation of emission
from physical processes such as accretion or star-formation 
 obscured by dust; however the precise nature of
the relative contribution of the two is still under study.
In the optical and ultraviolet, 
starlight is thought to dominate the total sky.  
Finally, at X-ray energies, the sky is thought to be dominated by emission 
from accreting supermassive black holes.  We refer to
the X-ray portion of the EBR as the cosmic X-ray background (CXB).

The spectrum of the CXB peaks
at high energies ($kT\approx 40$~keV), and the energy range over which
it may  be observed is quite large
 ($\approx0.1$~keV--1~MeV) representing a span of four decades in
 frequency.
  In the intervening years since the discovery of the CXB
(in the 1960's\cite{Giacconi62}), more
progress has generally been made towards resolving the background at
other wavelengths than in the X-ray band.  This is largely due to the
technical difficulty of building X-ray optics with spatial resolution
$\simlt 30^{\prime \prime}$.

The photon index of the CXB ($\Gamma=1.4$ from
1--10~keV\cite{miy98})
 is much harder than the $\Gamma\approx1.7$--2.3 typical of unobscured
AGN in the nearby Universe.  
 AGN synthesis models\cite{Setti89,Comastri95} can reproduce the CXB
spectrum from 1--20~keV but require the existence of a substantial obscured AGN
population (column densities in the range $N_{\rm
H}=10^{21}$--$10^{25}$~\cmsq).  The hardness of the CXB stands in
contrast to the types of AGN turned up in many of the deep X-ray
surveys prior to the \chandra--\xmm\ era (mainly Type I AGN), creating
the so-called ``spectral paradox." As will be described in the
following sections, this paradox is now partially solved.

\section{Science Results from Optical Follow-Up Programs Before
Chandra and XMM-Newton }
\subsection{The Soft ($< 2$~keV) Background }
\label{sect:softCXBgeneral}

A large fraction of the soft ($< 2$~keV) CXB was identified 
prior to \chandra\ and
\xmm\ through the combined sensitivity and imaging quality ($\simlt
5.0$\arcsec) of \rosat.   The deepest \rosat\ survey performed was the
UDS\cite{Lehmann01} of the Lockman Hole, chosen for its extremely low
Galactic absorption ($N_{H}\approx 5\times10^{19}$~cm$^{-2}$).  The UDS
sample includes 94 X-ray sources with $f_{\rm X} > 1.2 \times
10^{-15}$~\flux (0.5--2.0~keV) that were detected in a 
1112~ks \rosat\ High Resolution Imager (HRI) observation.  
At this flux level $\approx70$--80\% of the X-ray background 
in the 0.5--2.0~keV band is resolved.  The HRI observations 
were critical for optical identifications as they as they 
produced positions accurate to $\approx 2^{\prime \prime}$.  

Ground-based optical observations identified $\approx90$\% of the
UDS sources\cite{Lehmann01}.  The optically brightest counterparts 
were observed with the 
5~m Hale telescope, but the majority of the sources required a larger
aperture. The 10~m Keck telescope was used for this program as early as
February~1995\cite{Schmidt98}, showing that the marriage between deep
X-ray surveys and large telescopes goes back to the beginning of the
era of 10~m class telescopes.

Of these identified UDS sources, Type I (generally unobscured) AGN
comprised $\approx61$\% of the sources and Type II (generally obscured)
AGN comprised $\approx14$\% .  The Type I AGN were detected up to 
$z=4.45$\cite{Schneider98}, enabling the X-ray QSO luminosity function
to be constructed over a large interval of cosmic
time\cite{MiyajiXLF}.  Interestingly, the number density of luminous
QSOs as measured in the X-ray band did not demonstrate the decline
seen above $z\sim2$ in the optical band\cite{Schmidt95}\footnote{The
decline in luminous QSO number density at high redshift
as measured in the optical band has been extended to higher redshift
recently by Fan et~al.\cite{FanQLF}.  This work confirms the decline.}  
One possible interpretation for this difference between X-ray and
optical selection of high-redshift QSOs is a difference
in the bolometric luminosities being sampled.  Bolometric luminosity
may be correlated with the mass of the accreting black hole,
indicating possible differences in AGN evolution based on 
black hole mass\cite{MiyajiXLF}.

The UDS Type II AGN were typically 
found at lower redshifts (and lower X-ray luminosities).
The median redshift of the Type II AGN in the UDS 
survey was $z \approx0.6$\cite{Lehmann01}.
Recall (\S \ref{sect:xraysurveys}) that AGN synthesis 
models of the CXB predict a high fraction of obscured
AGN at high redshift\cite{Gilli99}.  Note also that at $z>3$ 
the \rosat\ band samples energies $> 2$~keV; at these hard energies
X-ray photons can easily penetrate column densities as high as 
$N_{\rm H} \approx 10^{22}$~\cmsq (and higher columns at even higher redshift).
The observations thus presented
a puzzle in that the population of luminous, obscured AGN
predicted by some CXB synthesis models\cite{Gilli99}  were not being found.

One of the results of the \rosat\ UDS that is also 
relevant to this contribution was the confirmation that
AGN tend to exist in a certain range of X-ray-to-optical flux ratio
($f_{X}/f_{V}$).  The X-ray to optical flux ratio had been found to be
a useful discriminator of object type in the
\einstein\ Medium-Sensitivity Survey (EMSS\cite{Stocke91}), where AGN
were noted to have $-1.0 < \log{f_{X}/f_{V}} < 1.2$ whereas normal
galaxies were found to have values of $\log{f_{X}/f_{V}} < -1.0$.
Schmidt et~al.\cite{Schmidt98} extended this to the greater depths of
the \rosat\ UDS, showing that 97\% of the UDS AGN had
$\log{f_{X}/f_{V}} > -1.0$.

The second most abundant class of objects in soft X-ray surveys (after
AGN) are clusters of galaxies.  There were 10 extended objects (the
majority of which were groups of galaxies) detected in the UDS over the
$\sim 30^{\prime}$ \rosat\ field of view\cite{Lehmann01}.  The clusters
included one which is double-peaked in the X-ray image and is possibly
two merging clusters of galaxies ($z=1.26$,
RXJ105343+5735\cite{Hasingerlensingcluster,Thompson01}).

Thus, the basic picture of the soft CXB prior to \chandra\ was that it
was mostly resolved, and most of the extragalactic X-ray  point sources
were Type I AGN.  The Type II AGN were found preferentially at lower
redshifts and lower X-ray luminosities.

\subsection{The Hard ($> 2$~keV) Background}
 \label{sect:hardCXBgeneral}

The greatest progress in understanding the CXB at $E > 2$~keV before
\chandra\ and \xmm\
 was made with the \asca\ (e.g., Akiyama et~al.\cite{Akiyama00}) and
\sax\ (e.g., Fiore et~al.\cite{Fiore01}) satellites.  Due to space
constraints I have focused on soft-band surveys;  this section is
in no way a comprehensive review of hard-band surveys.
Before these missions, only $\sim3$\% of the \hbox{$> 2$~keV} background was
resolved into individual point sources by \heao\cite{Picc82}.
\asca\ and \sax\ were able to identify the hard CXB sources to 2--10~keV fluxes
of $\approx 1\times10^{-13}$~\flux.  The expected range of
X-ray-to-optical flux ratios for AGN, calculated at this limit,
resulted in expected optical magnitudes of \hbox{$V\approx16$--21} for
the counterparts.  The positional accuracy of \asca\cite{Tanaka94} and
\sax\cite{Boella97} are comparable, typically $\approx30^{\prime
\prime}$.  In some cases, observations with other X-ray observatories
at softer energies allowed more precise positioning, but generally
multiple objects had to be observed spectroscopically to determine the
correct optical counterpart.  \asca\ and \sax\ were able to resolve and
identify $\approx20$--30\% of the hard CXB to 2--10~keV fluxes of
$\approx 1\times10^{-13}$~\flux.  At this level the number counts
($\log N$--$\log S$, $N$ is the number of sources per unit solid angle
brighter than a given flux, $S$) in the hard band are still consistent
with a Euclidean slope\cite{Akiyama00}.
A significant fraction of the hard CXB, although not the majority, 
was thus resolved before \chandra\ and \xmm\ and while AGN dominated
this resolved fraction, there still was not a large population of
high-redshift obscured AGN found.

   \begin{figure}
   \begin{center}
   \begin{tabular}{c}
   \includegraphics[height=11cm,angle=270]{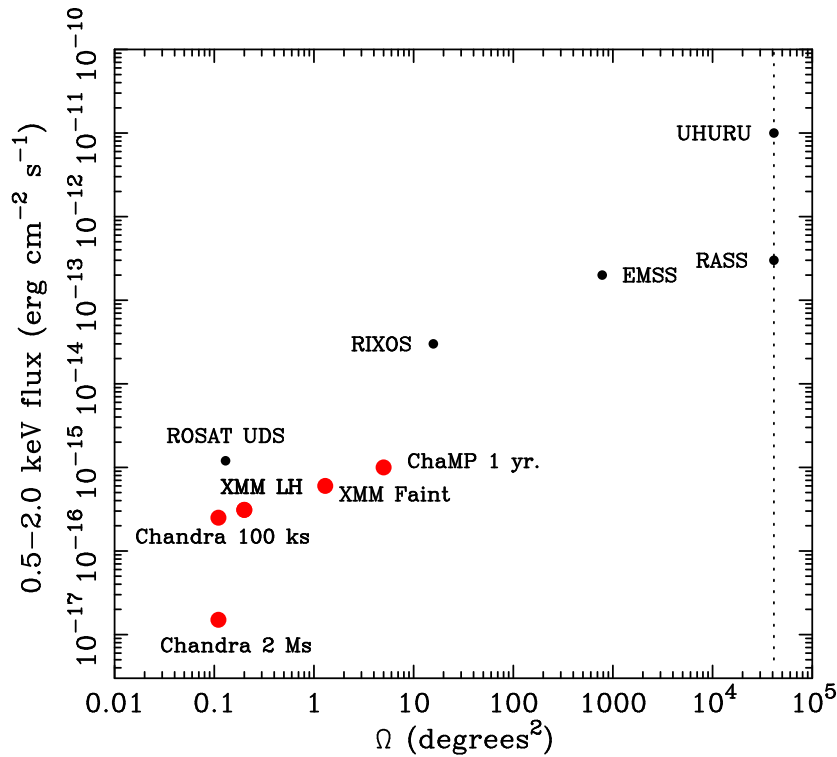}
   \end{tabular}
   \end{center}
   \caption[example]
   {Some extragalactic X-ray surveys in the 0.5--2~keV band, displayed
as limiting sensitivity versus solid angle.  The dashed vertical line
indicates coverage of the full sky.  The larger circles mark
current \chandra\ and \xmm\ surveys.
  Note that in 100~ks of observation with \chandra\, one is already
able to go significantly deeper than the previous generation of
X-ray surveys.  This figure is adapted from Brandt et~al.\cite{BrandtCatalog};
references for these surveys are given there.
 \label{fig:xrsurveysbrandt}
}
\end{figure}

   \begin{figure}
   \begin{center}
   \begin{tabular}{c}
   \includegraphics[height=10cm]{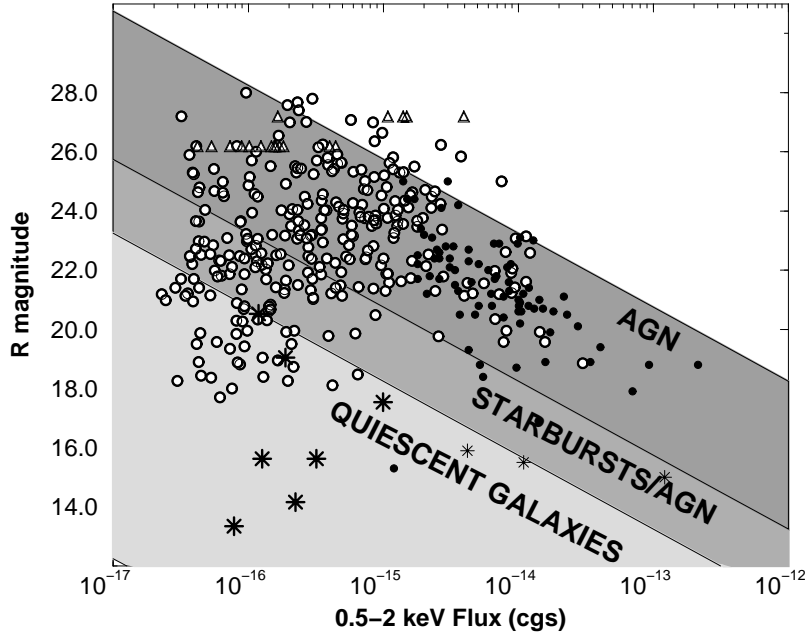}
   \end{tabular}
   \end{center}
   \caption[example]
   { \label{fig:fxfR}
$R$-band magnitude versus soft-band X-ray flux for Chandra Deep Field-North
sources within the high exposure area,
from Hornschemeier et~al.\cite{HornOBXF}; the CDF-N sources
are marked with open symbols.  The triangles mark upper limits, these
sources are potential $z>6$ AGN. 
  For comparison, the filled circles show the X-ray sources detected in
the \rosat\ UDS\cite{Lehmann01}.  
The shaded regions indicate typical values of X-ray-to-optical flux ratio
for different classes of objects. The sources located in the lowest 
 shaded region have X-ray-to-optical flux ratios well below those expected for
AGN; this is the regime of the normal and starburst galaxies.}
\end{figure}

\section{SCIENCE RESULTS FROM OPTICAL FOLLOW-UP OF X-RAY SOURCES}

\chandra\ has allowed soft band 
X-ray fluxes as low as $1 \times 10^{-16}$~erg~cm$^{-2}$~s$^{-1}$
and hard band X-ray fluxes as low as 
$6 \times 10^{-16}$~erg~cm$^{-2}$~s$^{-1}$ to become attainable with
moderately deep ($\approx 200$~ks) observations.  These fluxes are
$\approx10\times$ fainter in the soft band and $\approx100\times$ 
fainter in the hard band than the previous generation of deep surveys.
  \chandra's sub-arcsecond spatial resolution also allows 
for the first unambiguous matching to optical counterparts at hard
energies.   Some example
  \chandra\ and \xmm\ surveys are compared with other X-ray surveys in 
Figure~\ref{fig:xrsurveysbrandt}.  Note that this figure is just a representative
sample as the X-ray surveys carried out so far are too numerous to plot.

The science covered here includes results from the \chandra\ Deep Field (CDF) Surveys,
which have reached 2~Ms of coverage in the 
North\cite{BrandtCatalog,BargerCatalog2002,AlexanderCatalog} (hereafter CDF-N) 
and 1~Ms of coverage in the South\cite{Giacconi02} (hereafter CDF-S), 
representing the deepest X-ray views 
of the Universe to date.  Among the \xmm\ surveys, the Lockman Hole 
continues to be an important field for X-ray surveys\cite{HasingerXMM}\footnote{ 
Space prohibits additional discussion of the results from the Lockman Hole
survey, covered in \S\ref{sect:softCXBgeneral}. The reader is referred to 
Hasinger et al.\cite{HasingerXMM} for more discussion.}
Also discussed are some wide-field X-ray surveys, including the 
5 deg$^{2}$ \chandra\ Multiwavelength Project\cite{Wilkes01}
(ChaMP) and the ``An \xmm\ International Survey' (AXIS\cite{Watson01,Barcons02}).
Results are included from the 100~ks \chandra\ Spectroscopic 
Photometric Infrared-Chosen Extragalactic Survey
(SPICES\cite{SternSPICES}) survey.  Briefly discussed also is 
X-ray follow-up of sources discovered in optical surveys such as
the Sloan Digital Sky Survey (SDSS).

For the overall nature of optical follow-up of X-ray point 
sources\footnote{Diffuse 
X-ray sources are not discussed here; these are
clusters and groups of galaxies.  The
reader is referred to Bauer et~al.\cite{Bauer02} and references
therein for discussion of extended X-ray sources.}, 
refer to the plot of $R$-band magnitude versus 0.5--2~keV 
X-ray flux shown in Figure~\ref{fig:fxfR} for the CDF-N survey.  The range
of X-ray-to-optical flux ratio typical of AGN is
marked, and we see that it continues to hold for a large number of sources
down to (extremely!) faint optical magnitudes.  Attention should
be drawn to the faintest limits in both optical flux and X-ray flux. 
 At very faint optical fluxes (upper left of diagram), 
the sources are still consistent with the expected values for 
AGN, and this is where we expect high redshift (possibly even $z>6$) 
AGN to exist.  At very faint X-ray fluxes (left side), 
we find quite a few sources whose optical counterparts 
are {\it brighter} than expected for AGN--these sources are fairly 
``normal" galaxies, with their X-ray emission
dominated by processes associated with, among others, active star-formation 
and accreting binary systems.  We are now able to detect these normal galaxies 
in the X-ray band at cosmologically interestesting distances (i.e. $z > 0.1$).

\subsection{X-ray Selected AGN at High Redshift ($z > 3$)}
\label{sect:highzquasars}

Many of the X-ray sources detected in deep \chandra\ and \xmm\ 
surveys are optically faint ($I \simgt 24$)\cite{davofaint}.
This presents a real challenge to ground-based optical observatories, 
and exceeds the practical limit for optical 
spectroscopy even with the largest (10~m) telescopes.
The properties of the optically faint X-ray population 
(X-ray hardness, X-ray-to-optical flux ratio)
indicate that many are luminous obscured AGN at 
$z > 1$\cite{Fabian00,davofaint,BargerMushy,Gandhi02,Koeke02}, 
meaning that these sources provide an important window into the evolution
of obscured accretion power in the earlier Universe and may contain 
the elusive population of luminous obscured AGN at high redshift.

\begin{figure}
   \begin{center}
   \begin{tabular}{c}
   \includegraphics[width=15cm]{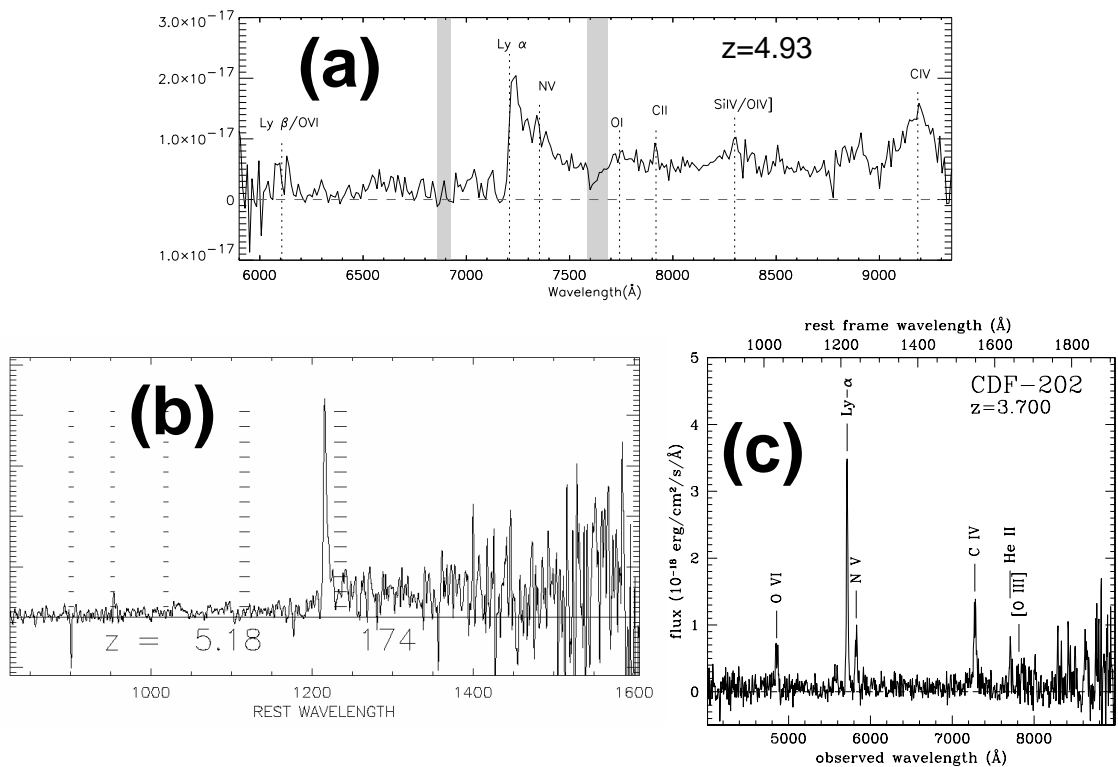}
   \end{tabular}
   \end{center}
   \caption[example]
   { \label{fig:highz}
A collection of optical spectra of high-redshift X-ray-selected AGN.  
(a) The New Technology Telescope (NTT) spectrum of a $z=4.93$ AGN
discovered serendipitously as part of the ChaMP survey,
adopted with permission from
Figure~2 of Silverman et~al.\cite{Silverman02}.
(b) A Keck LRIS spectrum\cite{BargerCatalog2002} of the $z=5.18$
X-ray selected AGN in the CDF-N.
(c) A VLT spectrum of the high-redshift Type 2 luminous AGN candidate
detected in the CDF-S.  The spectrum is adopted with
permission from Figure~2 of Norman et al.\cite{Norman02}.}

\end{figure}

Already there has been remarkable success in {\it selecting} AGN at very high
redshifts ($z \simgt 5$) with \chandra.
At the time of writing two very high redshift AGN have been selected in
deep X-ray surveys;
$z=4.93$ (discovered through ChaMP\cite{Silverman02}) and $z=5.18$
(discovered in the CDF-N\cite{BargerCatalog2002,Vignali02}).
The optical spectra of these
two highest redshift X-ray selected AGN are shown in Figure~\ref{fig:highz}.
\chandra\ is also able to detect Seyfert-luminosity objects at $z>4$, 
as evidenced by the detection of a lower-luminosity AGN at $z=4.424$ 
in the CDF-N\cite{BrandtIV}.

It is still the case that 
the vast majority of known $z >4$ AGN\footnote{There are $\approx 400$ 
$z>4$ AGN as of mid-2002.
http://www.astro.caltech.edu/$\sim$george/z4.qsos} have been discovered 
in the optical band.  This is largely an effect of the relative size of
 the areas surveyed.
For example, if we could reach the X-ray depth of 
the CDF surveys over the SDSS early data release area\cite{Stoughton02}
 ($\approx500$~deg$^{2}$), we would expect to find $\approx15$,$000$ AGN at $z >4$\cite{Vignali02}. 
There are $\approx45$ optically-selected 
AGN at $z>4$ in this area\cite{Stoughton02}.
The highest redshift quasars known to 
date were discovered in the SDSS\cite{Fanhighz} ($z=5.74$, 5.82,
5.99, 6.28) and all four of these have been detected 
with \chandra\ or \xmm \cite{Brandthighz}.  
Interestingly, despite the large change in the number density of optically
luminous QSOs\cite{Schmidt95,FanQLF},
there has not been strong evolution in certain properties, including
the optical-to-X-ray  spectral index, $\alpha_{\rm ox}$, in optically
selected (radio-quiet) AGN from $z\approx6$ to the present.

Optical selection of quasars of course carries a bias
against heavily absorbed objects that is less significant
in the X-ray band, in particular for high-redshift objects where
the observed X-ray emission comes from hard-to-absorb high-energy
photons.   It has been found that X-ray selected quasars
at $z>4$ have flatter values of 
$\alpha_{\rm ox}$\cite{Vignali02} (are more X-ray
bright) than optically selected $z>4$ quasars, 
as expected from optical selection effects. Among the optically
selected, X-ray faint $z>4$ AGN\cite{Vignali01} are those which tend to 
 exhibit signs of absorption, including strong ultraviolet absorption
features\cite{Gallagher02}. \xmm\ has found some strange absorbed AGN: 
 $\approx10$\% of the AXIS sources exhibit strong ultraviolet absorption but not
photoelectric absorption in the X-ray\cite{Barcons02}.  It is clear
that X-rays are an efficient means for selecting rare AGN populations which
will provide critical information on the nature of obscured activity
in the Universe\cite{Fabian99}.   

Pursuing the nature of ultraviolet absorption may 
be the key to understanding the early growth of
supermassive black holes. Are the material outflows which
possibly give rise to these absorption features much stronger 
at the higher accretion rates which may be present in nascent
black holes? Pursuing questions of this type requires sensitive near-infrared
spectroscopic observations.  An excellent example of this is
the observation of the $z=5.74$ \xmm-detected QSO by
two groups\cite{Maiolino01,Goodrich01}.  This object has $K\approx17$
and a high signal-to-noise spectrum was obtained with NIRSPEC in
2.7 hours of observation\cite{Goodrich01}.  If we are to
unambiguously identify all of the optically faint \chandra\ 
population (many of which are Extremely Red Objects with
$I-K > 4$), high signal-to-noise spectroscopy will be needed 
at $K\simgt24$.  This is clearly in the realm of 
large-aperture (30--100~m) telescopes.

As mentioned in \S \ref{sect:hardCXBgeneral}, X-ray background
synthesis models required a substantial population\cite{Wilman00}
 of highly obscured luminous AGN at hard X-ray energies.  
Sources with the expected hard X-ray spectra have been found in
deep X-ray exposures\cite{MushyNature,Horn01,Tozzi01}, 
but the spectroscopic identification of
these potentially high-redshift obscured AGN is still far 
from complete (both due to optical faintness and to the large size
of the areas surveyed).
There have been a few instances of successful
optical spectroscopic observations of interesting obscured AGN.
The \chandra\ Deep Field-South team has found
one convincing case for a luminous, high-redshift Type 2 QSO\cite{Norman02} 
for which they were able to obtain
a VLT optical spectrum (see Figure~\ref{fig:highz}) and another
object was found in the SPICES survey with Keck \cite{Stern02}.  
Perhaps there are
more such objects lurking in the unidentified X-ray source population, but
it already appears that the X-ray background synthesis models may need
major revision\cite{Fran02}, as a significant fraction of the hard
X-ray background flux is identified with sources
at fairly low redshift\cite{Rosati02,BargerCatalog2002}.  Of course,
the final answer awaits the positive identification of the remaining
$\approx 30$\% of the hard XRB flux density (and possibly for the higher
energy capability of future X-ray missions, see \S \ref{sect:nexgen}).

Ultradeep ($\simgt 1$~Ms) \chandra\ surveys may have already achieved 
the sensitivity to detect of the {\it first} supermassive black 
holes (SMBHs) to form
in the Universe at $z\approx8$--20.  X-ray surveys represent one
of the few ways that such objects might be found.  These SMBHs are
thought to play a crucial role in galaxy formation.
Such $\approx 10^{5}$--$10^{7}$~\msun
``proto-quasars" are expected to have X-ray luminosities comparable
to those of local Seyfert galaxies.  

   \begin{figure}
   \begin{center}
   \begin{tabular}{c}
   \includegraphics[height=10cm]{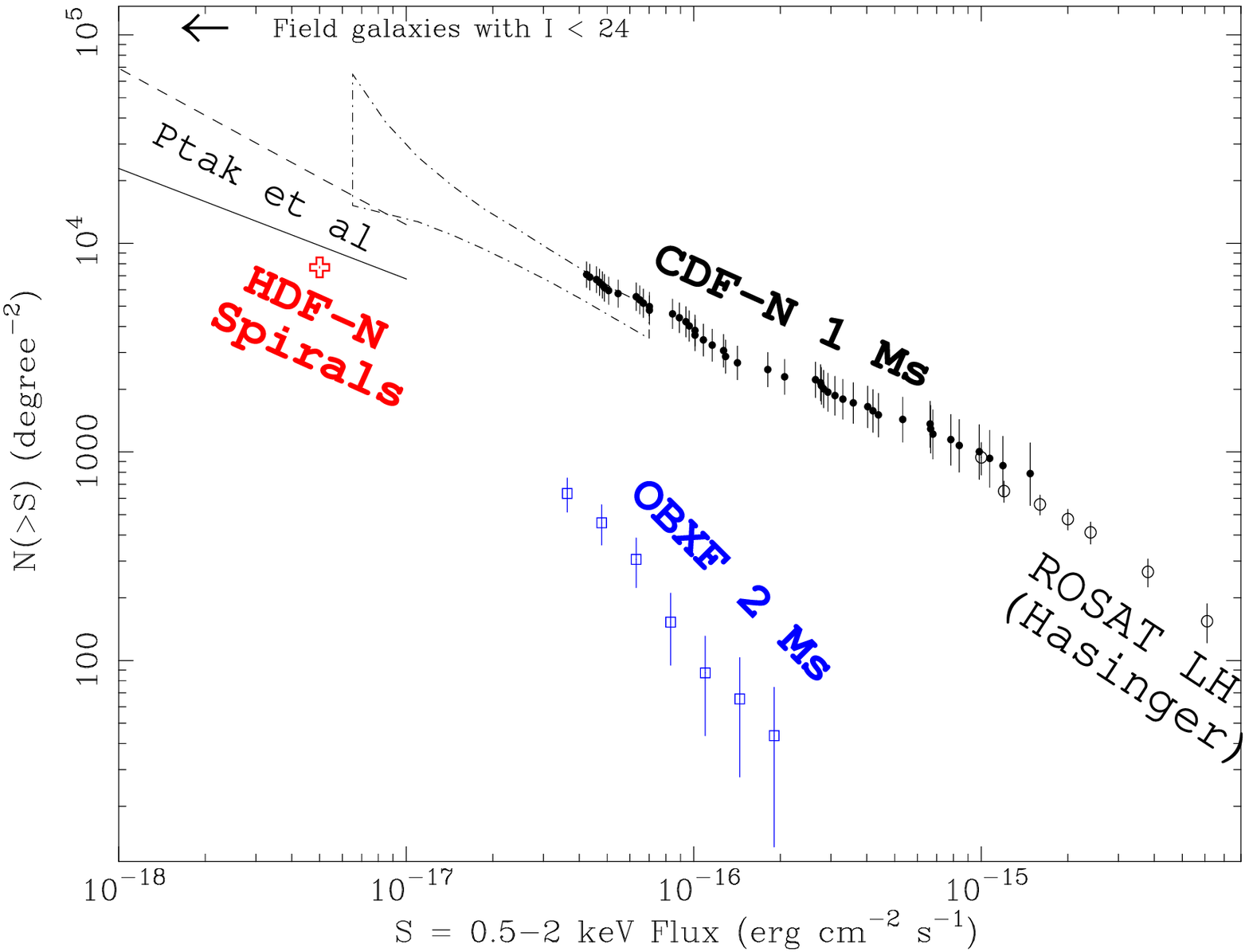}
   \end{tabular}
   \end{center}
   \caption[Number Counts]
   { \label{fig:LogNLogS}
Number counts for the X-ray detected ``normal" galaxies in the CDF-N 2~Ms survey
 as compared to other studies.  The black filled circles are the CDF-N 1~Ms data.\cite{BrandtCatalog}
The open squares indicate \rosat\ UDS data\cite{Hasinger98}.
The dashed and solid black lines at faint X-ray fluxes show two
predictions of the galaxy number counts from Ptak et~al.\cite{Ptak01}.
The open cross marks the constraint from the 1~Ms stacking
analysis of individually undetected field spirals\cite{Hornstacking}.
The arrow indicates the number density of field galaxies at $I=24$.
This figure adapted from Hornschemeier et~al.\cite{HornOBXF} with permission.

}
   \end{figure}

   \begin{figure}
   \begin{center}
   \begin{tabular}{c}
   \includegraphics[height=9cm]{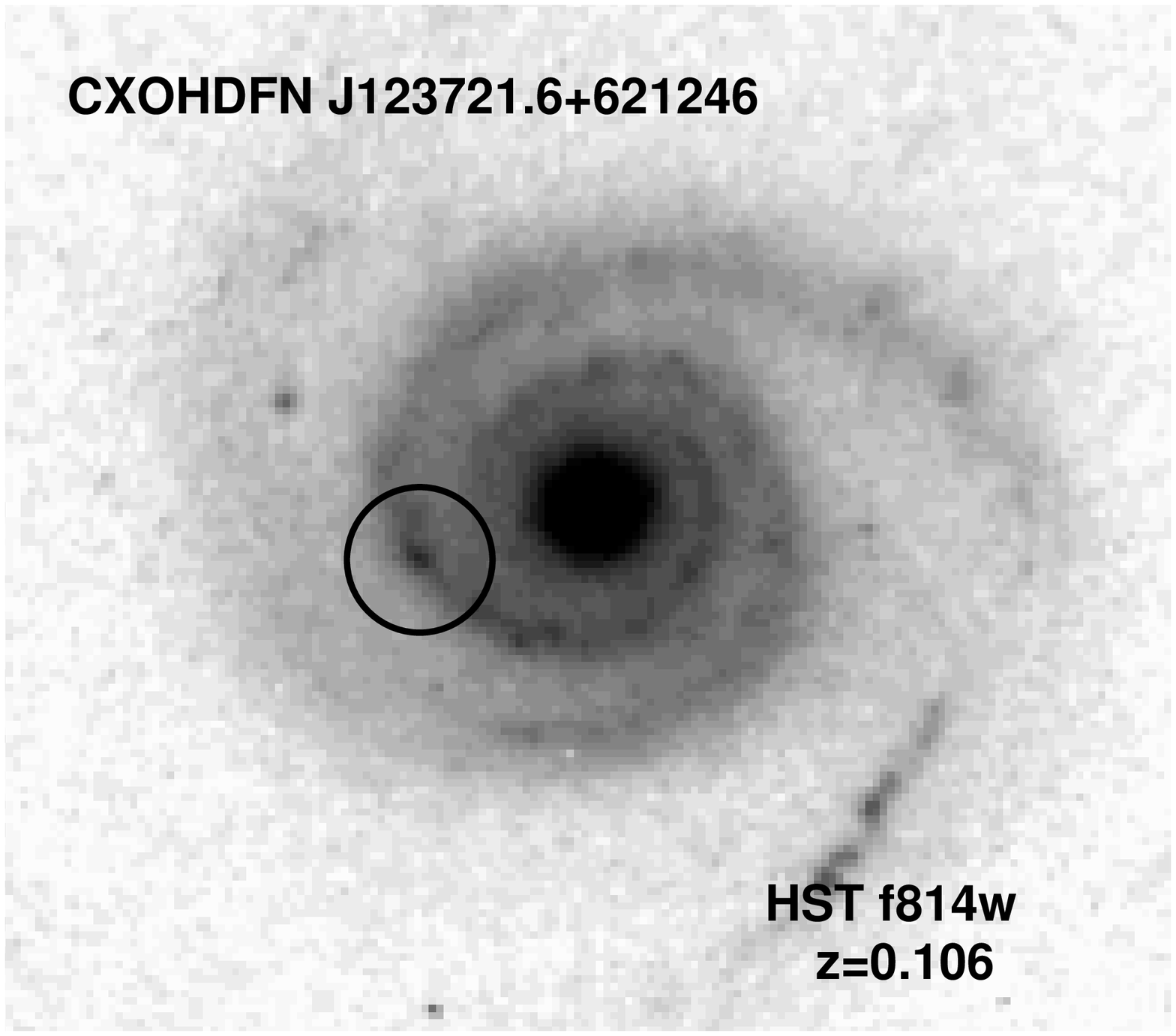}
   \end{tabular}
   \end{center}
   \caption[Off-Nuclear X-ray Source]
   { \label{fig:ULX}
HST f814w image  of a ULX within a galaxy in the CDF-N, 
adapted from Hornschemeier et al.\cite{HornOBXF}.
The location of the X-ray emission is marked by a circle of radius
1\farcs0, somewhat larger than the expected X-ray positional error.
The X-ray source appears to be located
along a spiral arm and is coincident with a region of slightly enhanced
optical emission. From X-ray variability analysis\cite{HornOBXF}, this
source is determined to be a black hole candidate.
}
   \end{figure}

\subsection{Starburst/normal Galaxies at Moderate Redshift ($z \simgt 1$)}

While many of the sources with the high X-ray-to-optical flux ratios
typical of AGN are extremely optically faint, other populations 
are arising in deep X-ray surveys
which are well-matched to the capabilities of 10~m telescopes.
One of these is the normal and starburst galaxy population,
whose X-ray emission is thought to originate from 
accreting binary star systems, 
supernova remnants, and the hot ISM.
What is new in the \chandra\ and \xmm\ era is that we are 
detecting normal galaxies at cosmologically interesting distances 
(look-back times of billions of years).  

Prior to \chandra\ and \xmm\ studies of galaxies
in the X-ray band had not reached far beyond $z\approx0.05$\cite{Fabbiano92}.  
In the CDF surveys, 
galaxies with  $0.1 \simlt z  \simlt 1.0$ 
are detected in appreciable numbers at 0.5--2~keV fluxes below 
$1 \times 10^{-15}$ erg~cm$^{-2}$~s$^{-1}$ (see
Figure~\ref{fig:fxfR}).  The bulk of the energy density of the 
CXB is certainly explained by AGN (see the preceding sections),  
but the investigation of the ``typical" galaxy, whether
its X-ray emission is dominated by  a population of X-ray binaries, hot
interstellar gas, or even a low-luminosity AGN, is an equally important
function of deep X-ray surveys.   Normal galaxies are likely to be the
most numerous extragalactic X-ray sources in the Universe and are
expected to dominate the number counts\cite{Ptak01} at 0.5--2~keV fluxes below  
$\approx 1 \times 10^{-17}$--$1 \times 10^{-18}$ erg~cm$^{-2}$~s$^{-1}$.  
It is only with \chandra\ that it has been possible to measure
the  normal and starburst galaxy X-ray number counts.
The number counts of the X-ray-detected normal galaxy population 
are much steeper\cite{HornOBXF} than that of the rest of the X-ray sources
discovered in deep surveys (see Figure~\ref{fig:LogNLogS}), and 
to resolve the last few percent of the X-ray background requires the presence
of this large population\cite{miy98}.

Reaching larger look-back times at high energies
presents  the exciting possibility of detecting the bulk X-ray response 
to the heightened star-formation rate at $z\approx1.5$--3\cite{Madau96}.
One expects the X-ray luminosity per unit 
$B$-band luminosity to be larger at $z\approx0.5$--1 
due to the increased production of X-ray binary progenitors
at $z\approx1.5$--3; this X-ray
emission represents a ``fossil record" of past epochs of star 
formation\cite{Ghosh01,Ptak01}. 
Therefore, measurements of the X-ray luminosities of typical galaxies 
can constrain models of X-ray binary production in galaxies.

Even with 1~Ms of X-ray coverage, many normal/starburst galaxies are
individually undetected in the X-ray band\cite{Hornstacking}.  
X-ray stacking analyses, possible because of the extremely low \chandra\
background,  allow for the study of these
individually undetected objects.  One such \chandra\ study (to $z\approx1$)
 has found\cite{Hornstacking} that 
 the average spiral galaxy is detected at 0.5--2~keV X-ray fluxes of 
$\approx$(3--5)$\times 10^{-18}$~erg~cm$^{-2}$~s$^{-1}$.
The X-ray to $B$-band luminosity ratio is found to be 
relatively constant to $z\approx 1$\cite{BrandtIV,Hornstacking}, 
but some upwards evolution is detected
by $z\approx 2$.  Since different global star-formation rates can lead to 
very different X-ray luminosity evolution profiles (e.g. Ghosh \& White\cite{Ghosh01}),
these constraints on the evolution of galaxies in the X-ray band are a useful
independent probe of the cosmic star-formation history.   

At higher redshift still ($z\approx$~2--4), the Lyman break technique 
has been used extensively to isolate actively star-forming 
galaxies\cite{Steidel96,Lowenthal97} and hence observe galaxies  
 near the peak of the cosmic star-formation rate\cite{Blain99}.
Lyman break galaxies have been stacked in the X-ray to find
an average rest-frame 2--8~keV luminosity 
of \hbox{$\approx 3.2\times 10^{41}$~erg~s$^{-1}$},
comparable to that of the most X-ray luminous starbursts in the local 
Universe\cite{BrandtLyBreak,Nandra02}.  The observed ratio of X-ray to $B$-band 
luminosity for these higher redshift galaxies is consistent with 
that seen from local starbursts. 
This technique of statistical X-ray analysis to survey objects
selected in the optical may also be useful at intermediate redshifts
to determine the properties of $z\approx1$ ``Balmer-Break" 
galaxies.  The Balmer-Break galaxies are found
 to have X-ray luminosities approximately five times lower than the Lyman-Break
galaxies\cite{Nandra02}, but the connection between the X-ray emission and the
ultraviolet emission of these galaxies is unclear.  What {\it is} clear
is that there is a new body of information on the evolution of
star-formation that remains to be fully exploited.

Deep X-ray surveys have also reached the sensitivity to detect individual
non-nuclear point sources {\it within} galaxies to appreciable 
distances ($z\approx0.3$). These sources are Ultra-luminous X-ray  
(ULX) sources with luminosities tens to hundreds of times higher than expected 
for Eddington-limited accretion onto a stellar-mass black hole (see Figure~\ref{fig:ULX}).  ULX sources represent either an intermediate-mass
 ($\approx 100$--500~${\rm M}_{\odot}$) class of black holes, or a 
beamed, relatively brief phase of more normal stellar mass black hole
 binary evolution\cite{King01}.  Intermediate-mass black holes are
of particular interest as they may be the seeds for the growth of
supermassive black holes.
Several candidate ULX sources within nearby 
galaxies have already been detected in the CDF-N, and continued follow-up
of X-ray surveys promises many more.  

Deep X-ray surveys with \chandra\ and \xmm\ cover roughly square
fields of side length \hbox{$\sim18$--30} arcminutes, requiring 
the wide-field capabilities of instruments such
as the VLT's Virmos (see Giampaolo Vettolani's contribution in this
same proceedings) and the Keck's DEIMOS (see Marc Davis' 
contribution in this same proceedings).  The required depth of coverage
to complement $\simgt 1$~Ms \chandra\ surveys is $R\approx23$ over
these large fields;  since many of these normal galaxies are 
absorption-dominated, 10~m telescopes are clearly required.

\section{The Next Generation of Observatories}
\label{sect:nexgen}

In the late 1990's, it was demonstrated that 10~m telescopes and
deep \rosat\ surveys were well matched; $\approx96$\% of the \rosat\
UDS sources were successfully identified though optical ground-based
observations\cite{Lehmann01}.   In the early 2000's, 
$\approx 30$\% of the sources detected in X-ray surveys 
with \chandra\ and \xmm\  are beyond the reach of optical spectroscopic
observations with 10~m optical telescopes but populations such as 
X-ray detected normal and starburst galaxies are found to be well-matched
with 10~m telescopes.  Within the optically
faint population lies the
answer to the question of how much of the accretion activity
of the Universe is obscured and key parameters concerning
the growth of the first supermassive black holes of the Universe. 
Within the X-ray detected galaxy population is crucial information
on the growth of stellar and intermediate-mass black holes and new
physical measures of star-formation processes. 
The identification of many of the optically faint 
sources must necessarily wait for the light-collecting
power of 30--100~m telescopes.   The
normal and starburst galaxies will be well-studied for the 
first time now that wide-field multi-object spectrographs are
available on 10~m class telescopes (e.g., VLT Virmos and Keck
DEIMOS).

What does the next generation of X-ray telescopes have in store?
An excellent description of NASA's future missions in X-ray
astronomy is given in White et~al.\cite{White02} and in a review
of new X-ray missions is presented in SPIE conference 4835 by
Richard Mushotzky. The reader
is also referred to the proceedings of the SPIE conference on 
X-ray missions (4851).   Here I give a brief summary of what is to come
in the next few decades.

The \conx\ mission, which is planned for launch around 2010,
will provide X-ray {\it spectroscopy} down to the faint 
limits of the CDF surveys, where formerly we have relied upon 
broad-band X-ray photometry.  Recall that the collecting area of \chandra's
mirrors is $\approx100$--600~cm$^{2}$ over 0.5--8~keV (the collecting
area of \xmm\ is higher at $\approx400$--1300~~cm$^{2}$ over 0.5--10~keV). 
 \conx\ is planned to have 15,000~~cm$^{2}$ effective area
at 1~keV and spectroscopic coverage
up to 40~keV \footnote{The reader is referred to the \conx\ web page
for the latest details on the mission design, http://constellation.gsfc.nasa.gov/}. This large collecting area will allow for detailed emission-line
analysis, we will be able to physically constrain the nature of
the material near accreting supermassive black holes at $z\approx8$ and
resolve the emission lines
in normal and starburst galaxies up to $z\approx1$.  X-ray spectroscopy
of individual stellar-mass black holes up to $z\approx0.3$ will also
become possible.  

The \xeus\ and \genx\  observatories are currently being planned on 
longer timescales (10--20~years) to study X-ray sources down to 
0.5--2~keV fluxes of $\approx 4 \times 10^{-18}$~erg~cm$^{-2}$~s$^{-1}$ and 
$\approx 5 \times 10^{-20}$~erg~cm$^{-2}$~s$^{-1}$, respectively.
It can be expected that the optical counterparts to many of these
X-ray sources will be extremely faint ($R\simgt30$ if the trend of
X-ray-to-optical flux ratio continues to these X-ray fluxes!).  At
this level, every quiescent and starburst galaxy in the Universe
will be detected in the X-ray band to at least $z\approx3$ 
and deep X-ray surveys
shall closely resemble the optical Hubble Deep Field observations
as far as areal density of sources.  So, in addition to depth, 
high quality imaging will be needed, possibly only attainable with
space-based near-infrared imagers such as \ngst.
It is clear that X-ray astronomy will continue to present 
many challenges for future generations of optical telescopes.

\appendix    

\acknowledgments     

The author thanks the organizers of the ``Discoveries and
Research Prospects from 6-to-10~m Telescopes" for the invitation
to speak; with special thanks to the session chair 
Puragra Guhathakurta.
The author gratefully acknowledges the financial support of
NASA grant NAS~8-38252 (Gordon P. Garmire, PI), 
CXC grant G02-3187A (W.N. Brandt, P.I.), and  
NASA GSRP grant NGT5-50247 throughout the past several years.
The author thanks the \chandra\ X-ray Observatory  and 
\chandra\ Deep Field-North teams for making much of the work
presented here possible.  She especially thanks the \chandra\
X-ray Center for her upcoming \chandra\ Fellowship.
The author thanks Sarah Gallagher, David Alexander, Niel Brandt
and Donald Schneider for reading a draft of this proceedings 
prior to submission and Guenther Hasinger, Colin Norman, 
and John Silverman for providing figures.

%


\bibliography{reportA}   
\bibliographystyle{spiebib}   

\end{document}